\documentstyle[12pt,epsf]{article}

\def\hybrid{\topmargin -20pt    \oddsidemargin 0pt
        \headheight 0pt \headsep 0pt
        \textwidth 6.25in       % A4 paper
        \textheight 9.5in       % A4 paper
        \marginparwidth .875in
        \parskip 5pt plus 1pt   \jot = 1.5ex}

\hybrid

\def\baselinestretch{1.2}

\catcode`\@=11

\def\marginnote#1{}
\newcount\hour
\newcount\minute
\newtoks\amorpm
\hour=\time\divide\hour by60
\minute=\time{\multiply\hour by60 \global\advance\minute by-\hour}
\edef\standardtime{{\ifnum\hour<12 \global\amorpm={am}%
        \else\global\amorpm={pm}\advance\hour by-12 \fi
        \ifnum\hour=0 \hour=12 \fi
        \number\hour:\ifnum\minute<10 0\fi\number\minute\the\amorpm}}
\edef\militarytime{\number\hour:\ifnum\minute<10 0\fi\number\minute}
\def\draftlabel#1{{\@bsphack\if@filesw {\let\thepage\relax
   \xdef\@gtempa{\write\@auxout{\string
      \newlabel{#1}{{\@currentlabel}{\thepage}}}}}\@gtempa
   \if@nobreak \ifvmode\nobreak\fi\fi\fi\@esphack}
        \gdef\@eqnlabel{#1}}
\def\@eqnlabel{}
\def\@vacuum{}
\def\draftmarginnote#1{\marginpar{\raggedright\scriptsize\tt#1}}

\def\draft{\oddsidemargin -.5truein
        \def\@oddfoot{\sl preliminary draft \hfil
        \rm\thepage\hfil\sl\today\quad\militarytime}
        \let\@evenfoot\@oddfoot \overfullrule 3pt
        \let\label=\draftlabel
        \let\marginnote=\draftmarginnote
   \def\@eqnnum{(\theequation)\rlap{\kern\marginparsep\tt\@eqnlabel}%
\global\let\@eqnlabel\@vacuum}  }

\def\preprint{\twocolumn\sloppy\flushbottom\parindent 2em
        \leftmargini 2em\leftmarginv .5em\leftmarginvi .5em
        \oddsidemargin -.5in    \evensidemargin -.5in
        \columnsep .4in \footheight 0pt
        \textwidth 10.in        \topmargin  -.4in
        \headheight 12pt \topskip .4in
        \textheight 6.9in \footskip 0pt
        \def\@oddhead{\thepage\hfil\addtocounter{page}{1}\thepage}
        \let\@evenhead\@oddhead \def\@oddfoot{} \def\@evenfoot{} }

\def\numberbysection{\@addtoreset{equation}{section}
        \def\theequation{\thesection.\arabic{equation}}}

\def\underline#1{\relax\ifmmode\@@underline#1\else
        $\@@underline{\hbox{#1}}$\relax\fi}

\def\titlepage{\@restonecolfalse\if@twocolumn\@restonecoltrue\onecolumn
     \else \newpage \fi \thispagestyle{empty}\c@page\z@
        \def\thefootnote{\fnsymbol{footnote}} }

\def\endtitlepage{\if@restonecol\twocolumn \else \newpage \fi
        \def\thefootnote{\arabic{footnote}}
        \setcounter{footnote}{0}}  %\c@footnote\z@ }

\catcode`@=12
\relax

\def\figcap{\section*{Figure Captions\markboth
        {FIGURECAPTIONS}{FIGURECAPTIONS}}\list
        {Figure \arabic{enumi}:\hfill}{\settowidth\labelwidth{Figure
999:}
        \leftmargin\labelwidth
        \advance\leftmargin\labelsep\usecounter{enumi}}}
 \relax
\def\tablecap{\section*{Table Captions\markboth
        {TABLECAPTIONS}{TABLECAPTIONS}}\list
        {Table \arabic{enumi}:\hfill}{\settowidth\labelwidth{Table
999:}
        \leftmargin\labelwidth
        \advance\leftmargin\labelsep\usecounter{enumi}}}
 \relax
\def\reflist{\section*{References\markboth
        {REFLIST}{REFLIST}}\list
        {[\arabic{enumi}]\hfill}{\settowidth\labelwidth{[999]}
        \leftmargin\labelwidth
        \advance\leftmargin\labelsep\usecounter{enumi}}}
 \relax
\makeatletter
\newcounter{pubctr}
\def\publist{\@ifnextchar[{\@publist}{\@@publist}}
\def\@publist[#1]{\list
        {[\arabic{pubctr}]\hfill}{\settowidth\labelwidth{[999]}
        \leftmargin\labelwidth
        \advance\leftmargin\labelsep
        \@nmbrlisttrue\def\@listctr{pubctr}
        \setcounter{pubctr}{#1}\addtocounter{pubctr}{-1}}}
\def\@@publist{\list
        {[\arabic{pubctr}]\hfill}{\settowidth\labelwidth{[999]}
        \leftmargin\labelwidth
        \advance\leftmargin\labelsep
        \@nmbrlisttrue\def\@listctr{pubctr}}}
 \relax
\makeatother
\newskip\humongous \humongous=0pt plus 1000pt minus 1000pt

\newif\ifdtup

\relax

\def\be{\begin{equation}}
\def\ee{\end{equation}}
\def\ba{\begin{eqnarray}}
\def\ea{\end{eqnarray}}

\def\del{\partial}

\def\a{\alpha}

\def\b{\beta}

\def\g{\gamma}
\def\G{\Gamma}

\def\D{\Delta}

\def\m{\mu}
\def\n{\nu}

\def\L{\Lambda}
\def\s{\sigma}

\def\no{\noindent}

\def\qq{\qquad}

\def\IR{\relax{\rm I\kern-.18em R}}

\def \ha {{1\over 2}}

\def \ov {\over}

\def\IR{\relax{\rm I\kern-.18em R}}
\def\inv{^{\raise.15ex\hbox{${\scriptscriptstyle -}$}\kern-.05em 1}}

\def\tL{{\tilde L}}

\begin{document}
%\draft

\renewcommand{\theequation}{\arabic{equation}}

\newcommand{\beq}{\begin{equation}}
\newcommand{\eeq}[1]{\label{#1}\end{equation}}
\newcommand{\ber}{\begin{eqnarray}}
\newcommand{\eer}[1]{\label{#1}\end{eqnarray}}
\newcommand{\eqn}[1]{(\ref{#1})}
\begin{titlepage}
\begin{center}

%Phys. Lett. {\bf B432} (1998) 365 
\hfill CERN-TH/99-63\\
\hfill hep--th/9903109\\

\vskip .8in

{\large \bf On Asymptotic Freedom and Confinement \\from Type-IIB Supergravity }

\vskip 0.6in

{\bf A. Kehagias}\phantom{x} and\phantom{x} {\bf K. Sfetsos}
\vskip 0.1in
{\em Theory Division, CERN\\
     CH-1211 Geneva 23, Switzerland\\
{\tt kehagias,sfetsos@mail.cern.ch}}\\
\vskip .2in

\end{center}

\vskip .6in

\centerline{\bf Abstract }

\no
We present a new type-IIB supergravity vacuum that
describes the strong coupling regime of a non-supersymmetric gauge theory.
The latter has a running coupling such that the theory becomes 
asymptotically free in the ultraviolet. It also has a running   
theta angle due to a non-vanishing axion field in the 
supergravity solution.
We also present a worm-hole solution, which
has finite action per unit four-dimensional volume and  
two asymptotic regions, a flat space and an $AdS^5\times S^5$.
The corresponding ${\cal N}=2$ gauge theory, instead of being 
finite, has a running coupling. We compute the quark--antiquark potential 
in this case and find that it exhibits, under certain assumptions,
an area-law behaviour for large separations.

\vskip 0,2cm
\no

\vskip 4cm
\noindent
CERN-TH/99-63\\
March 1999\\
\end{titlepage}
\vfill
\eject

\def\baselinestretch{1.2}
\baselineskip 16 pt
\noindent

%%%%%%%%%%%%%%%%%%%%%A generalization of target space%%%%%%%%%
\def\tT{{\tilde T}}
\def\tg{{\tilde g}}
\def\tL{{\tilde L}}

%%%%%%%%%%%%%%%%%%%

\section{Introduction}

One of the well-known vacua of type-IIB supergravity is the $AdS_5\times S^5$
one, first described in \cite{SS}. The non-vanishing fields here are the
metric and a Freund--Rubin type  anti-self-dual five-form. This background has
received a lot of attention recently because of its conjectured connection to
${\cal N}=4$ $SU(N)$ supersymmetric Yang--Mills (SYM) theory at large $N$
\cite{malda,kleb}.
According to this conjecture \cite{malda},
the large-$N$ limit of certain superconformal field theories (SCFT)
can be described in terms of Anti de-Sitter (AdS)
supergravity; correlation functions of the SCFT theory that lives 
in the boundary
of AdS can be expressed in terms of the bulk theory \cite{kleb}.
In particular,  the four-dimensional ${\cal N}=4$ 
$SU(N)$ SYM theory is described by the type-IIB string theory  on
$AdS_5\times S^5$, where the radius of both the $AdS_5$ and  
$S^5$ are proportional to $N$. 

However, one is interested for obvious reasons 
in  non-supersymmetric YM theories.  
In particular, the existence of supergravity duals of such theories will help 
in understanding their strong-coupling behaviour.  
There are a number of proposals in this direction. 
One of them is within the type-II0 theories \cite{KT,Mh,E}, which, 
although non-supersymmetric, are consistent string 
theories \cite{DH}. 
These theories have a tachyon in their spectrum due to lack 
of supersymmetry. The tachyon is coupled to the other fields of the theory,
namely, the graviton, the dilaton and the antisymmetric-form fields. 
In particular, the coupling of the tachyon to the dilaton is such that it 
drives  the latter to smaller values in the ultraviolet (UV), a hint for 
asymptotic freedom. 

However, a different approach has been proposed  in 
\cite{KS}. According to this, {\it the supergravity duals of 
non-supersymmetric gauge theories are non-supersymmetric background solutions 
in type-IIB theory}. Based on this, a solution with a 
non-constant dilaton that corresponds to a  gauge theory 
with a UV-stable fixed point has been found \cite{KS}.
The coupling approaches the fixed point with a
power-law behaviour. The solution 
is valid for strong 't Hooft coupling $g^2_{\rm H}$, which is consistent 
with the fact that there are no known perturbative field theories with 
UV-stable fixed point. This scenario has also been followed 
in \cite{Gubs-GPPZ}.
The same power-law behaviour as in \cite{KS} has also been found in the 
case of an ${\cal N}=2$ boundary gauge theory from a supergravity vacuum
with both D3- and D7-branes \cite{demelo}.  

\section{A non-supersymmetric solution}

Here, we will try  
to describe a celebrated feature of gauge theories, namely the asymptotic 
freedom, within the type-IIB theory. As we will see, 
this is possible if we make a ``minimal'' extension 
to the solution we found in \cite{KS} by turning on the other scalar field 
of type-IIB, the axion.     
The anti-self-dual 
five-form $F_5$ is given by the Freund--Rubin-type ansatz, which is
explicitly written as
\ba
F_{\mu\nu\rho\kappa\lambda}&=& -\frac{\sqrt{\Lambda}}{2}
 \epsilon_{\mu\nu\rho\kappa\lambda}\ , \qq
\mu,\nu,\ldots=0,1,\dots,4\ , 
\nonumber\\
F_{ijkpq}&=& \frac{\sqrt{\Lambda}}{2} \epsilon_{ijkpq}\ , \qq
i,j,\dots=5,\dots,9\ ,
\label{FF}
\ea
and it is clearly anti-self-dual. 
We will also assume, for the metric, four-dimensional 
Poincar\'e invariance $ISO(1,3)$, 
since we would like a gauge theory defined on Minkowski space-time. 
In addition,
we will preserve the  $SO(6)$ symmetry of the supersymmetric 
$AdS_5\times S^5$ vacuum. 
As a result, the $ISO(1,3)\times SO(6)$ invariant ten-dimensional metric, 
in the Einstein frame, is of the form 
\be
ds^2=g_{\mu\nu}dx^\mu dx^\nu+  g_{ij}dx^idx^j\ ,
\label{anz0}
\ee
where    
\be 
g_{\mu\nu}dx^\mu dx^\nu =dr^2 + K(r)^2 dx_\a dx^\a\ , \qq \a=0,1,2,3\ ,
\label{anz1}
\ee
and $g_{ij}$ is the metric on $S^5$. The dilaton and the axion, by 
$ISO(1,3)\times SO(6)$ invariance, can only be a function of $r$.
In the Euclidean regime, the supergravity equations for the 
metric, the dilaton and the axion in the Einstein frame 
follow from the action \cite{GGP} 
\be
S=\frac{1}{\kappa^2}\int d^{10}x\sqrt{g}\left(R-\frac{1}{2}(\partial\Phi)^2
+\frac{1}{2}e^{2\Phi}(\partial\chi)^2\right)\ . 
\label{1action}
\ee
Note the minus sign in front of the axion kinetic term,
which is the result of the Hodge-duality rotation of the type-IIB 
nine-form  \cite{GGP}. Then the field equations, taking into account the 
anti-self-dual form as well, are
\ba
R_{MN}&=&\frac{1}{2}\partial_M\Phi\partial_N \Phi-\frac{1}{2}e^{2\Phi}
\partial_M\chi
\partial_N\chi+\frac{1}{6}F_{MKLPQ}{F_N}^{KLPQ}\ , 
\label{RRR}\\
\nabla^2\Phi&=&-e^{2\Phi}\left(\partial\chi\right)^2\ ,  
\label{akjs}\\
\nabla^2\chi& =& 0\ .
\label{Fi}
\ea
The equation of motion of the 
five-form  is the anti-self-duality condition, which is satisfied for the 
ansatz  \eqn{FF}. For the particular case we are considering here, the above 
equations turn out to be  
\ba
&&R_{\m\n} = - \L g_{\m\n} + \ha \del_\m \Phi\del_\n \Phi
 - \ha e^{2\Phi} \del_\m \chi\del_\n \chi\ ,
\label{RRR1} \\
&& {1\ov \sqrt{-g} } \del_\m \left(\sqrt{-g} g^{\m\n} \del_\n \Phi\right)
 =-e^{2\Phi}\partial_\n\chi\partial_\m\chi g^{\m\n} \ , \label{FFFF}\\
&& {1\ov \sqrt{-g} } \del_\m \left(\sqrt{-g} g^{\m\n} e^{2\Phi}
\del_\n \chi\right)
 =0\ ,
\label{eqs1}
\ea
and 
\be
R_{ij} \ =\  \L g_{ij} \ . \label{eqs12}
\ee
Equation \eqn{eqs12} is automatically solved for a five-sphere of radius 
$2/\sqrt{\L}$, and  a first integral of the axion in eq. \eqn{eqs1} is
\ba
 \chi' = \chi_0 K^{-4}e^{-2\phi}\ ,
\label{eqs2}
\ea 
where the prime denotes derivation with respect to $r$, 
and $\chi_0$ is a dimensionless integration constant. 
With no loss of generality, it can be taken to be positive by appropriately 
changing the sign of $r$.
Using this expression for $\chi$ in \eqn{FFFF} we obtain the differential 
equation 
\be
K^4(K^4\Phi')'=-e^{-2\Phi}\chi_0^2\ ,  \label{FFFFF}
\ee
a first integral of which is given by 
\be
K^8\Phi'^2=\chi_0^2e^{-2\Phi} + \mu\ . \label{fFFF}
\ee
Equations \eqn{eqs2} and \eqn{fFFF} are sufficient to proceed and solve
for the function $K(r)$ that appears in the metric \eqn{anz1}. 
The non-zero components
of the Ricci tensor for the metric \eqn{anz1} are 
\ba
R_{rr} & = & - 4 \frac{K''}{K^2} \ ,
\nonumber\\
R_{\a\b} & = & - \eta_{\a\b} \left( KK''
+ 3 K'^2 \right)\ ,
\label{ricc1}
\ea
and \eqn{RRR1} is equivalent to the following differential equation
\ba
&&K'^2= {\mu\ov 24}\ K^{-6} + {\L\ov 4}\ K^2\ .
\label{eqs3}
\ea
The solution of the above equation 
for $\mu=\alpha^2>0$, as a function of $r$, is
\be
K^4=\frac{\alpha}{\sqrt{6\Lambda}}\sinh\left(2 \sqrt{\Lambda}
(r_0-r)\right) \ ,\qq r\leq r_0\ ,
\label{k44}
\ee
where $r_0$ is an integration constant. 
It should be noted that the case $\mu<0$ gives rise to a dilaton with bad 
asymptotics  and it will not be considered here. 
Substituting the expression in \eqn{k44} in \eqn{fFFF} and \eqn{eqs2},
we find that the string coupling and the axion 
are given by\footnote{The computation is greatly facilitated if 
we first change variables as 
$$z=\int K^{-4}dr=\sqrt{\frac{3}{2 \a^2}}
\ln \coth\left(\sqrt{\L} (r_0-r)\right)\ .$$}
\ba
&&e^\Phi= {\chi_0\ov 2 \a} 
\left( \Big( \coth \sqrt{\L} (r_0-r)\Big)^{\sqrt{3}/2}
-\Big( \tanh(\sqrt{\L} (r_0-r)\Big)^{\sqrt{3}/2} \right) \ ,
\label{eff1}\\
&&\chi=-{\a\ov \chi_0}\
 {\Big(\coth \sqrt{\L} (r_0-r)\Big)^{\sqrt{3}} + 1\ov 
\Big(\coth \sqrt{\L} (r_0-r)\Big)^{\sqrt{3}} - 1 }\ .
\label{eff2}
\ea

It is convenient to identify
the parameters $\a$ and $\Lambda$  
in such a way that, in the limit $r\to -\infty$,
the Einstein metric becomes that of $AdS_5 \times S^5$. This requires that
$\a= 4 \sqrt{6} R^3 e^{-4 r_0/R}$ and $\Lambda=4/R^2$. On the other hand,
$r$ cannot take arbitrarily large values since, at $r=r_0$, the function
$K(r)$ has a zero 
and the space is singular in both the Einstein and the string frames. 
The presence 
of this singularity makes the extrapolation from the UV region, 
where the space is $AdS_5\times S^5$, to the IR region problematic, which 
is ultimately related to the lack of supersymmetry. 
As we will see next, there are supersymmetric 
backgrounds that are non-singular in the sense that there are geodesically
complete.  

\section{A supersymmetric solution}

The aspect we would like to address here is supersymmetry. For this we 
need the 
fermionic variations \cite{GGP}, which are written, in our background as 
\ba
\delta \lambda &=&
%-e^{\Phi}\left(\partial_M\chi+\partial_Me^{\Phi}\right)
-\frac{1}{2}\left(e^{\Phi}\chi'-\Phi'\right)\gamma_r\epsilon^*\ , \nonumber \\
\delta\lambda^* & =&
 -\frac{1}{2}\left(e^{\Phi}\chi'+\Phi'\right)\g_r \epsilon\ ,
\label{lamb}\\
\delta\psi_M & =& \left(\nabla_M+\frac{1}{4}e^\Phi\partial_M\chi
-\frac{\sqrt{\Lambda}}{4}\gamma_M\right)\epsilon\ , \nonumber \\
\delta\psi_M^*& =& \left(\nabla_M-\frac{1}{4}e^\Phi\partial_M\chi
-\frac{\sqrt{\Lambda}}{4}\gamma_M\right)\epsilon^*\ , 
\label{dpsi}
\ea
where $\lambda,~\psi_M$ are the dilatino and the gravitino, respectively, 
and $\gamma_M$ are the $\gamma$-matrices.  
We may easily check, using  \eqn{fFFF}, that the background is supersymmetric 
for 
\be
\mu=0\ , ~~~~~~\chi'=\pm e^{-\Phi}\Phi'\  .
\label{muu}
\ee
In that case, there exist Killing spinors $\epsilon=e^{\pm \Phi/4}\zeta$ 
where $\zeta$ are Killing spinors on the (Euclidean) 
$AdS^5\times S^5$. By choosing 
$\chi'= e^{-\Phi}\Phi'$,  the background  
breaks half of the supersymmetries  and we find that 
\ba
K(r)& =& e^{-r/R_0}\ , \\
e^\Phi& =& g_s+\frac{R_0}{4}\chi_0e^{4r/R_0}\ . \label{PP}\\
\chi& =&a_\infty -e^{-\Phi}\ , \label{xxx}
\ea
where $a_\infty$ is the constant value of the axion field at infinity and
$R_0^4=4 \pi N$.
After changing coordinates as $\rho=R_0e^{r/R_0}$, 
the metric for the supersymmetric solution
in the string frame becomes
\ba
ds^2&=&\left(1+\frac{\chi_0}{4R^3 g_s^{1/4}}\rho^4\right)^{1/2}
\left(\frac{R^2}{\rho^2}\right.\left(d\rho^2+dx^\alpha dx_\alpha\right)
+R^2d\Omega_5^2\left.\!\!\!\!\!\!\!\phantom{\frac{1}{\rho^2}}\right)\\
&=&R^2\left(\frac{\chi_0}{4R^3 g_s^{1/4}}\right)^{1/2}
\left(1+\frac{4g_s^{1/4}R^3}{\chi_0}\frac{1}{\rho^4}
\right)^{1/2}\left(d\rho^2+\rho^2 d\Omega_5^2+dx^\a dx_\a\right)\ ,
\ea
where $R^4=g_s R_0^4=4 \pi g_s N$.
Thus, the background is conformally flat and it has two asymptotic
regions
\ba
&&\rho\rightarrow \infty ~,~~~~~~\mbox{flat space}\ , \\
&&\rho\rightarrow 0 ~, ~~~~~~~~\mbox{$AdS_5\times S^5$ \ .}
\label{INT}
\ea
The action of this vacuum is easily found, following \cite{GGP}, to be
\be
S=-\frac{1}{2\kappa^2}\int d^{10}x \sqrt{g}\nabla^2\Phi=\frac{c}{g_s}\ ,~~~~~~
c=\frac{1}{2\kappa^2}\chi_0R_0^5V(S^5)\int d^4x \  , \label{ACT}
\ee
where $V(S^5)=\pi^3$.
Thus, it has a finite value per unit 
four-dimensional volume. In addition it has the standard $1/g_s$ scaling of
D-branes. 
As a result, this solution can be interpreted as an interpolating soliton 
between flat space and Euclidean $AdS_5\times S^5$. 
We also note that this solution corresponds to the large size limit of the
D-instanton solutions of \cite{BGKR,SHW}. 

Concerning the above solution, we would like to stress that 
it breaks half of the supersymmetries even in the two asymptotic 
regions, where the space is either flat or $AdS_5\times S^5$. 
Thus, the corresponding boundary gauge theory will be 
an ${\cal N}=2$ SYM theory, which, however, will 
have running coupling, as we will see. Hence, it should be  
different from the one obtained by 
orbifolding the ${\cal N}=4$ theory \cite{KSs}.

\section{Running couplings and the\\
 quark-antiquark potential}

In order to discuss the running of couplings, we must identify the coordinate
in the bulk that corresponds to the energy in the boundary gauge theory.
We identify 
the energy in the bulk as $U= R^2 e^{-r/R}$ and also define an energy scale 
$U_1= R^2 e^{-r_0/R}$, which, since $U>U_1$, may be considered as the 
IR cutoff. The identification of the
energy  follows from the fact \cite{SW} that if one considers the massless
scalar equation 
${1\ov \sqrt{G}}\del_M e^{-2 \Phi} \sqrt{G} G^{MN}\del_N \Psi=0$, 
where $G_{MN}$ is the $\s$-model metric, this takes the form of the usual 
scalar equation in $AdS_5 \times S^5$ to leading order in the expansion for
large $U$. According to the AdS/CFT correspondence, the dependence of the 
bulk fields on $U$ may be interpreted as energy dependence of the boundary 
theory such that  long (short) distances in the AdS space corresponds to 
low (high) energies in the CFT side. In particular, the $U$-dependence of the 
dilaton defines the energy dependence of the YM coupling 
$g_{\rm YM}^2=4\pi e^{\Phi}$, as well as of the 't Hooft coupling $g_{\rm H}^2=
g_{\rm YM}^2N$. 
According to this we find the running of the 't Hooft coupling to be
\be
g_H = g_0 \Big({U_1\ov U}\Big)^4 + {\cal O}\Big({U_1\ov U}\Big)^{12} \ ,
\label{betrt}
\ee
where $g_0=\sqrt{3} \chi_0/R^4 \a$.
As we see, $g_{\rm H}$ vanishes asymptotically, the signal 
of asymptotic freedom. It should be stressed, however, that the 
't Hooft coupling approaches zero with a power law and not with the familiar 
perturbative logarithmic way. Note also that $\b_g'(0)= -4$, in accordance 
with the universal behaviour for the first derivative of the beta function 
for the coupling that was found in \cite{KS}.
In the supersymmetric solution in particular, one finds, by using \eqn{PP},
that the beta function is 
\be
\beta(g_{\rm H})=-4(g_{\rm H}-g^*_{\rm H})\ , \label{hoo}
\ee
where $g_{\rm H}^{*2}=4\pi N g_s$. 

Next we turn to the quark--antiquark potential 
corresponding to the supersymmetric solution of section 2, by 
computing the Wilson loop.
Using standard techniques \cite{Mal1,Rey} we find that the potential is
\be
E_{q\bar q} = \Big(-U_0  + {a \ov U_0^3}\Big) {\eta_1 \ov \pi} \ ,\qq
a={\chi_0 R^5\ov 4 g_s^{1/4}}\ ,
\label{qbqrq}
\ee
where $\eta_1 ={\pi^{1/2} \Gamma(3/4)\ov \G(1/4)}\simeq 0.599$,
%$T$ is the total range of the Euclidean time variable, 
and $U_0$ is the turning point of the trajectory given by 
\ba
U_0 &=& a^{1/4} \left( {1\ov 6} \D^{1/3} - 2 \D^{-1/3}\right)^{-1/2}\ ,
\nonumber\\
\D & \equiv &  
108 b^2 + 12 \sqrt{12 + 81 b^2}\ ,\qq b={a^{1/4}\ov 2 \eta_1 R^2} L\ .
\label{gfgh}
\ea
It turns out that, for $a^{1/4}/R^2 L \ll 1$, the first term in \eqn{qbqrq}
is dominant, resulting in the usual Coulombic law behaviour \cite{Mal1,Rey}.
The first correction to it is
${\cal O}(L^3)$, as in \cite{KS,demelo}, which is also similar to 
the behaviour one finds \cite{BISY1} using near extremal D3-branes to describe
finite-temperature effects in the ${\cal N}=4$ $SU(N)$ SYM theory at large $N$.
However, in the opposite limit of $a^{1/4}/R^2 L\gg 1$, the second term
in \eqn{qbqrq} dominates, giving 
\be
E_{q\bar q} \simeq {a^{1/2}\ov 2\pi R^2}\  L\ ,
\label{coonf}
\ee
where we have used the fact that $U_0\simeq a^{1/4}/b^{1/3}$ in that limit.
Hence, the quark--antiquark potential exhibits the typical confining behaviour
and produces the area-law for the Wilson factor.
The plot of the potential \eqn{qbqrq} is given below.
%\vspace{2cm}
\begin{figure}[htb]
\epsfxsize=4in
\bigskip
\centerline{\epsffile{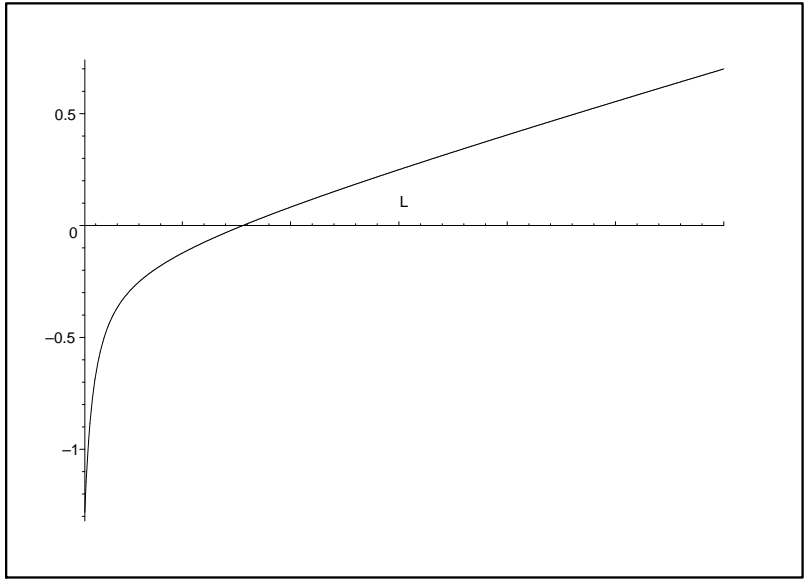}}
\caption{Plot of the quark--antiquark potential in \eqn{qbqrq} as a function of
the separation distance $L$, in units of $a=(2 \eta_1 R^2)^4$. 
For small $L$ we have a Coulombic behaviour, whereas it is linear for large 
$L$.}
\bigskip
\label{fig1}
\end{figure}

%\vskip -.6 cm
A second criterion \cite{witten} for confinement is the existence of 
a mass gap in the theory. In our case it is straightforward to check this,
by examining the massless wave equation and realizing that the
Einstein metric is just $AdS_5 \times S^5$.

At this point we emphasize that since we go into the strong coupling 
regime for $U=0$,
we should introduce an infrared cutoff $U_{\rm IR}$ such that 
$U\ge U_{\rm IR}$. 
This does not affect the behaviour in the UV, but it has consequences
in the IR. It implies that there is a maximum length $L_{\rm IR}$, up to 
which \eqn{coonf} can be trusted. Under this assumption, the range 
of validity of \eqn{coonf} is $a^{-1/4} R^2 \ll L \ll L_{\rm IR}$.
An estimate for $L_{\rm IR}$ is found as follows: It is natural to take for
$U_{\rm IR}$ the value for $U$ where the string coupling becomes of order 1,
i.e. $U_{\rm IR}\sim g_s^{1/4} a^{1/4}$.
Then the cutoff is compared with 
the minimum value for $U$, i.e. $U_0\sim U_{\rm IR}$, which 
implies that $L_{\rm IR}\simeq a^{-1/4} R^2 g_s^{-3/4}$. 
Since $g_s\to 0$ (keeping the constant $a$ in \eqn{qbqrq} finite)
we see that our result \eqn{qbqrq} is practically valid everywhere.

\bigskip\bigskip

\centerline{\bf Note added}

Just before submitting our paper to the hep-th we received \cite{tseliu},
which overlaps with material in section 3.

\end{document}